Foam stabilization in salt solutions : the role of capillary drainage and Marangoni stresses

Ekta Sharma [a#], Suraj Borkar [a#], Philipp Baumli[b], Xinfeng Shi[b], James Y.Wu[b], David Myung[ac], Gerald G. Fuller [a*]

a Department of Chemical Engineering, Stanford University, Stanford, California 94305; and b Alcon Research LLC, Fort Worth, Texas 76134, United States; c Spencer Center for Vision Research, Byers Eye Institute, Department of Ophthalmology, Stanford University School of Medicine, Palo Alto, California 94303

# Equal contributions

## Abstract

The long-standing question of why foaming is easier in seawater than in freshwater remains unresolved. In this study, we address this issue through precise interferometric single-bubble experiments, demonstrating that the theory proposed by G. Marrucci (1969) provides a compelling explanation. Electrolyte solutions with varying concentrations of phosphate salts were used to study film formation and drainage, with thickness tracked by interferometry. In deionized water, bubbles rupture within seconds due to rapid dimple collapse. However, in phosphate salt solutions, bubbles persisted for several minutes.

While surface tension gradients from evaporation-driven salt concentration gradients have been thought to create Marangoni stresses, our results show that despite film thinning being capillary drainage-dominated, Marangoni-driven influx can be observed. Marrucci's theory explains this by showing that an increased interfacial area as the film thins, leads to higher salt concentration in the film due to Gibbs surface excess. This concentration gradient induces Marangoni stresses, causing flow reversal, increased film thickness, and enhanced foam stability.

We show that Marrucci's theory has been incorrectly dismissed, and the predicted critical heights where fluid influx occurs closely match our findings and other studies using sodium chloride. Additionally, we extend the theory's applicability to foam films in non-aqueous fluid mixtures, highlighting its broader relevance.

## Introduction

Foaming is widespread and inherent in various liquid environments. The drainage dynamics of the fluid film sandwiched between two bubbles, also known as the foam film, govern the stability of the foam. Bubbles coalesce when the liquid film between them drains out, eventually leading to the collapse of the foam. It is well known that pure liquids do not foam, whereas foaming is commonly observed in ocean waves [1–3]. The salts present in seawater contribute to foam stability, but the exact mechanisms by which they stabilize the foam remain unclear despite numerous studies. Foams are also used in a wide range of industrial applications, including wastewater treatment, mineral processing, personal care products, pharmaceuticals, textiles, food processing, and pulp and paper industries [4–6].

Salts can suppress the drainage dynamics of the thin liquid film around a bubble through various mechanisms. These include electrostatic stabilization via increased electrostatic disjoining pressure [7], increased film viscosity [8], crystals getting trapped in the plateau borders [9], and Marangoni flow [10]. Whether salts suppress or promote foaming typically depends on ion-specific coalescence behavior, which arises from the synergistic effects of the ions present in the thin film, their segregation, and their molecular interactions with the liquid phase, which dictate foam stabilization [11]. Several studies have focused on identifying the critical concentrations of salts beyond which coalescence is inhibited [12–16].

While the stabilization of bubbles in the presence of surfactants is well understood, the role of salts in this stabilization process has received less attention [17]. Extensive research has explored the combined effects of salts, surfactants, and particles on the stabilization of bubbles and foams [9,18]. Surfactants reduce surface tension, while salts tend to increase it. However, the complex interplay between these factors is often overlooked in many studies, particularly regarding the role of surface tension variations. Surfactants adsorb at the air-water interface, rendering it immobile and making the thin film flow parabolic. When a bubble ascends in a surfactant solution and reaches the air-water interface, the surface tension in the thin fluid film around the bubble is typically higher than in the bulk. This creates an inward flow of fluid due to Marangoni stresses, thereby enhancing the bubble's stability against coalescence.

In contrast, salts increase surface tension by causing ions to either deplete from or accumulate at the air-water interface, leading to a rise in surface tension within the film region [19–24]. The increased surface tension drives an inward flow from the bulk to the thin fluid film, which can enhance the stability of the foam. Despite the significance of this process, the specific impact of salts in the absence of surfactants is still not well understood.

This work aims to elucidate the specific impact of salts on foam stabilization in the absence of surfactants. We investigate the stability of isolated bubbles in the presence of salts, explicitly examining their resistance to coalescence. The coalescence of an isolated bubble with ambient air mimics inter-bubble coalescence in a bulk liquid and coalescence at the bulk liquid-air interface. Interferometry is used to analyze the drainage dynamics of the foam film around a bubble in an electrolyte solution. Our observations suggest that the foam film drainage is primarily governed by capillary drainage, with minimum film thickness in the electrolyte solution $h_{\min} \sim t^{-2/3}$ [25–28]. However, we also observed an unexpected influx of liquid into the film region, which cannot be fully explained by existing theories based solely on Marangoni stresses from evaporation.

One of the pioneering theories in the study of film drainage in electrolyte solutions is that of G. Marrucci, who in 1969 proposed a mechanism that accounts for the influence of surface tension gradients on foam film stability [12]. Marrucci's theory suggests that as the film thins, the interfacial area increases, leading to a change in the bulk concentration of ions in the film relative to the bulk solution. This concentration gradient creates a surface tension gradient, which in turn induces a Marangoni flow that opposes further thinning of the film. Marrucci's theory provides a framework for understanding how electrolyte solutions can enhance foam stability, especially when the Péclet number—an indicator of the relative importance of advection versus diffusion in mass transport—is high. Despite its potential, Marrucci's theory has been largely overlooked in recent years, possibly due to a lack of precise experimental validation at the time.

Historically, experimental approaches to understand foam stability have been limited by the difficulty of accurately measuring the thickness of thin liquid films [11,29,30]. Interferometric techniques have become increasingly popular because they provide precise, real-time measurements of film thickness, allowing researchers to track the evolution of foam films under different conditions [31–39]. These techniques have enabled detailed studies of film drainage dynamics and have provided valuable insights into the role of electrolytes in foam stabilization.

In this study, we revisit Marrucci's theory using modern experimental techniques, specifically interferometric single-bubble experiments, to investigate the role of electrolyte concentration in foam film stability. Phosphate salts were selected as model electrolytes due to their strong impact on surface tension and their application as a food emulsifier [19,40]. We systematically vary the concentration of these salts and track the evolution of film thickness using interferometry. Our goal is to test whether Marrucci's theory can accurately predict the critical film thickness at which Marangoni flows reverse the drainage process, stabilizing the film. Additionally, we extend our analysis to non-aqueous systems, such as silicone oil blends, and lithium bromide (LiBr) dissolved in propylene carbonate, to assess the broader applicability of the theory [31,41–44]. Our results show that the film thickness decreases to a critical value, below which Marangoni stresses dominate over Laplace pressure, governing the drainage dynamics.

Finally, we compare our experimentally observed critical film thickness with theoretical estimates from Marrucci's work [12], finding that they are of the same order of magnitude (~1 µm). We also compare our results with those reported in studies by Liu et al. (2023) [45], Karakashev et al. (2008) [46], and Firouzi and Nguyen (2014) [47], which investigated NaCl solutions. In all cases, Marrucci's theory predicts critical heights that agree well with experimental observations, supporting its validity across different systems.

## Materials and methods

### Materials

Sodium phosphate dibasic anhydrous was obtained from Fischer Scientific, and potassium phosphate monobasic was sourced from Sigma-Aldrich. Deionized water with a resistivity of 18.2 MΩ·cm (Milli-Q) was used for all experiments. The phosphate salt solution was prepared by dissolving equal weight proportions (1:1 w/w) of sodium phosphate dibasic and potassium phosphate monobasic in deionized water, using a magnetic stirrer to ensure complete dissolution. The saturation concentration $\left(c_{sat}\right)$ of 7000 mol $m^{-3}$ was determined by lyophilizing a saturated salt solution.

**Methods**

The experimental procedure followed the methodology outlined by Frostad et al. [39], with modifications made to improve the quality of interference patterns [48] (see Figure 1). These adjustments involved optimizing the optical elements to enhance image clarity. Approximately 5 mL of the aqueous phase was introduced into the chamber, and an air bubble with a volume of 0.8 µL was formed using a syringe pump. The bubble volume was monitored by a side camera. Using live image processing, the bubble was held on a 16-gauge needle (ID: 1.19 mm, OD: 1.65 mm) such that its apex was positioned at a distance equal to the bubble radius from the air-water interface.

The air-water interface was then lowered toward the bubble (with the bubble stationary) using a motorized stage. The interface approached the bubble by a distance 1.1 times the bubble radius at an approach velocity of 150 µm $s^{-1}$. This velocity is significantly lower than the critical value $\sqrt{\sigma/2\rho R}$ ensuring slow viscous drainage rather than rapid inertial drainage [49,50].

The thin liquid film formed between the bubble and the air-water interface exhibited interferograms, which were captured using a 10x magnification lens and a camera. A white light source illuminated the thin film region. These interferograms were processed into film thickness values using custom software developed in Python with Qt [39]. Experiments were performed at room temperature, with a relative humidity of 48 ± 3%. The chamber was cleaned thoroughly between experiments by sequential washing with heptane, water, and acetone, followed by drying.

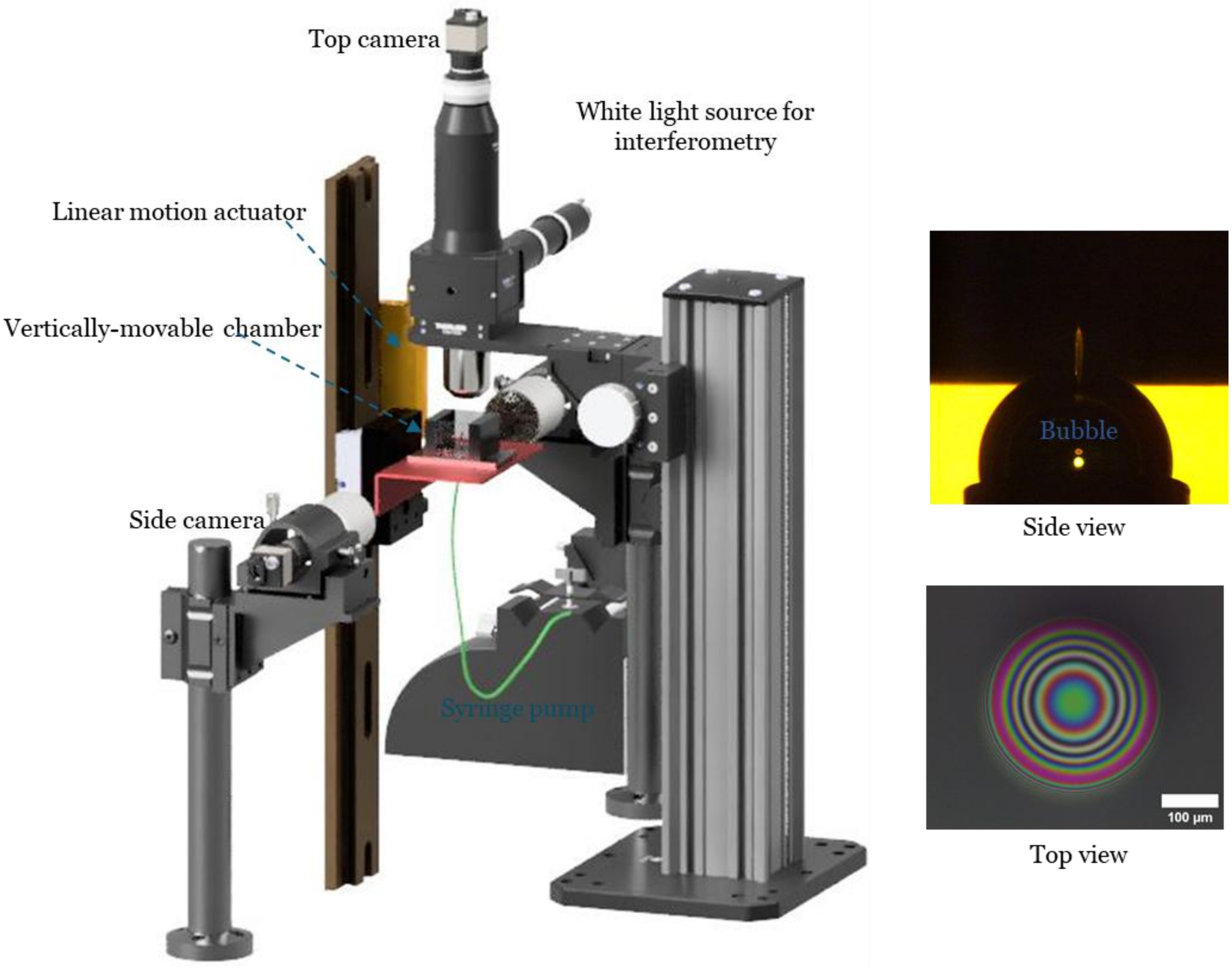

**Figure 1**: Schematic of the dynamic fluid-film interferometer setup, accompanied by typical images from the side and top cameras. A bubble is created at the tip of a stainless-steel needle immersed in an aqueous medium using a syringe pump, with the bubble size monitored and controlled via the side camera. The flat air-water interface is positioned at a distance from the bubble apex equal to the bubble's radius. The interface is then moved downward at a velocity of 150 μm $s^{-1}$ by a distance of 1.1 times the bubble radius. A collimated white light source is directed through a 10X objective lens, and the reflected rays from both the flat air-water interface and the bubble-water interface are collected by the same objective lens and directed to a color camera. These reflected rays interfere, producing an interference pattern that provides detailed information about the shape and dynamics of the foam film formed.

A bulk foam test was also conducted to examine the effect of salts on foam stabilization (see Figure 2). A fine glass frit funnel with a pore size of 40 μm and a volume of 500 mL was used. The funnel was filled with 100 mL of the aqueous phase, and airflow was maintained at 16 mL $s^{-1}$ for 2 minutes. Afterward, the airflow was stopped, and foam stability was observed.

Surface tension of deionized water and salt solutions was measured using the pendant drop method, and viscosity was measured with a rheometer (Discovery HR-3, TA Instruments) equipped with a cone and plate geometry (2° cone angle, 40 mm diameter).

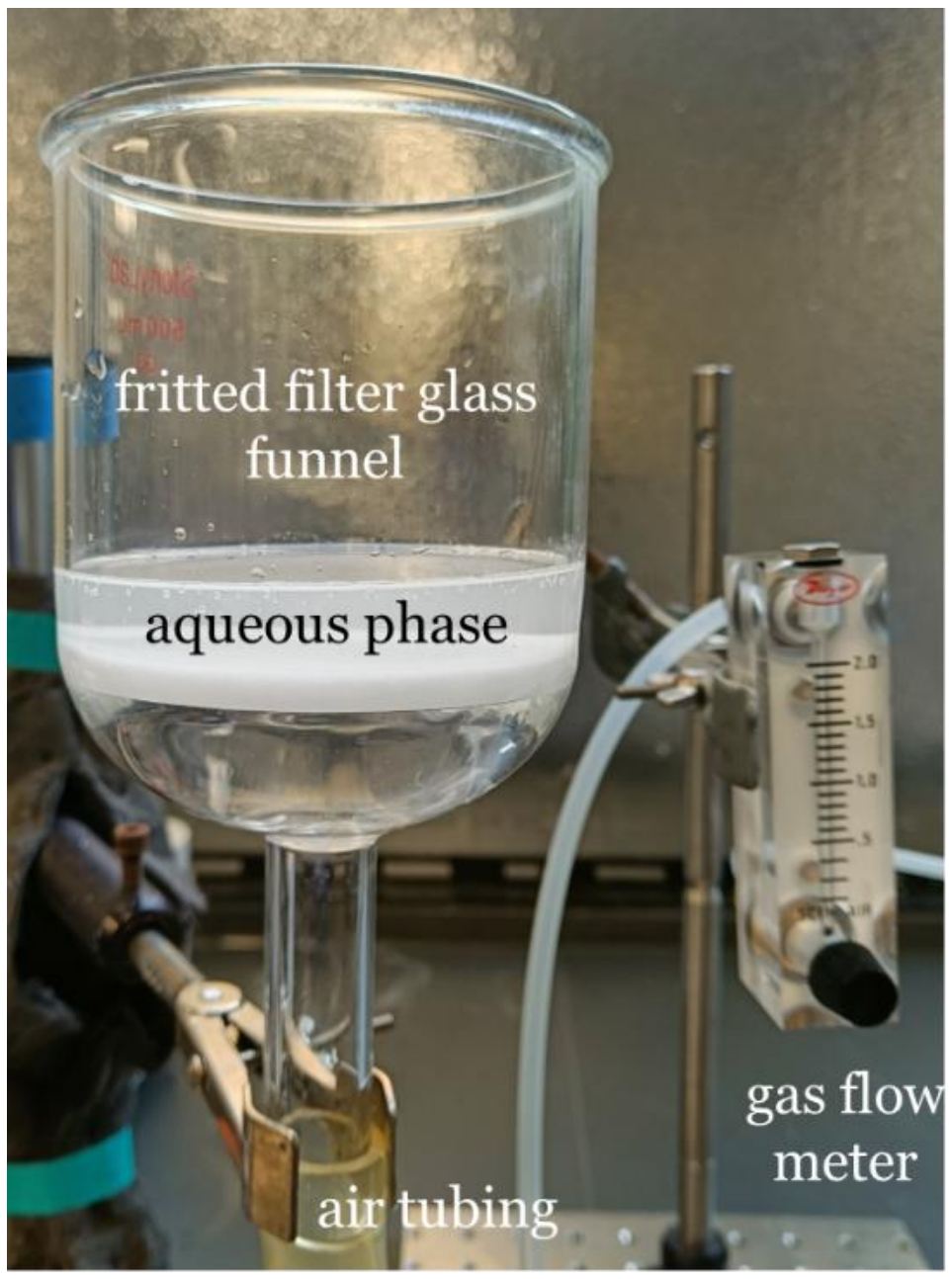

**Figure 2**: Experimental setup for observing bulk foam stability. A fritted glass funnel is filled with the aqueous medium, and air is sparged through the medium to generate foam. The stability and lifetime of the bubbles formed are then qualitatively analyzed.

## Results and discussion

### Surface tension and viscosity of electrolyte solutions

Before delving into film drainage dynamics, it is crucial to confirm that the surface tension and viscosity of concentrated phosphate solutions increase as previously reported [19]. Figure 3 presents the surface tension and viscosity of the salt solutions, formed by mixing equal proportions of sodium phosphate dibasic and potassium phosphate monobasic in water, as a function of salt concentration.

The surface tension, $\sigma$, increased linearly with salt concentration, $c$, with a slope of $6.34 \times 10^{-6}$ N m$^{2}$ mol$^{-1}$ (see Figure 3). Previous studies by Christenson & Yaminsky (1995) [51] and Weissenborn & Pugh (1995) [8] proposed that salts with $(d\sigma/dc)^2$ values on the order of 1 mN² m⁻² M⁻² or higher hinder bubble coalescence. However, exceptions to this trend have been documented, as reviewed by Craig (2004) [14]. In our study, phosphate salts were chosen specifically for their high $(d\sigma/dc)^2$ value of 41 mN² m⁻² M⁻², well above the threshold, offering deeper insight into the stabilization mechanism. A wider range of concentrations was explored to better understand the underlying processes, as suggested by Christenson et al. (2004) [52].

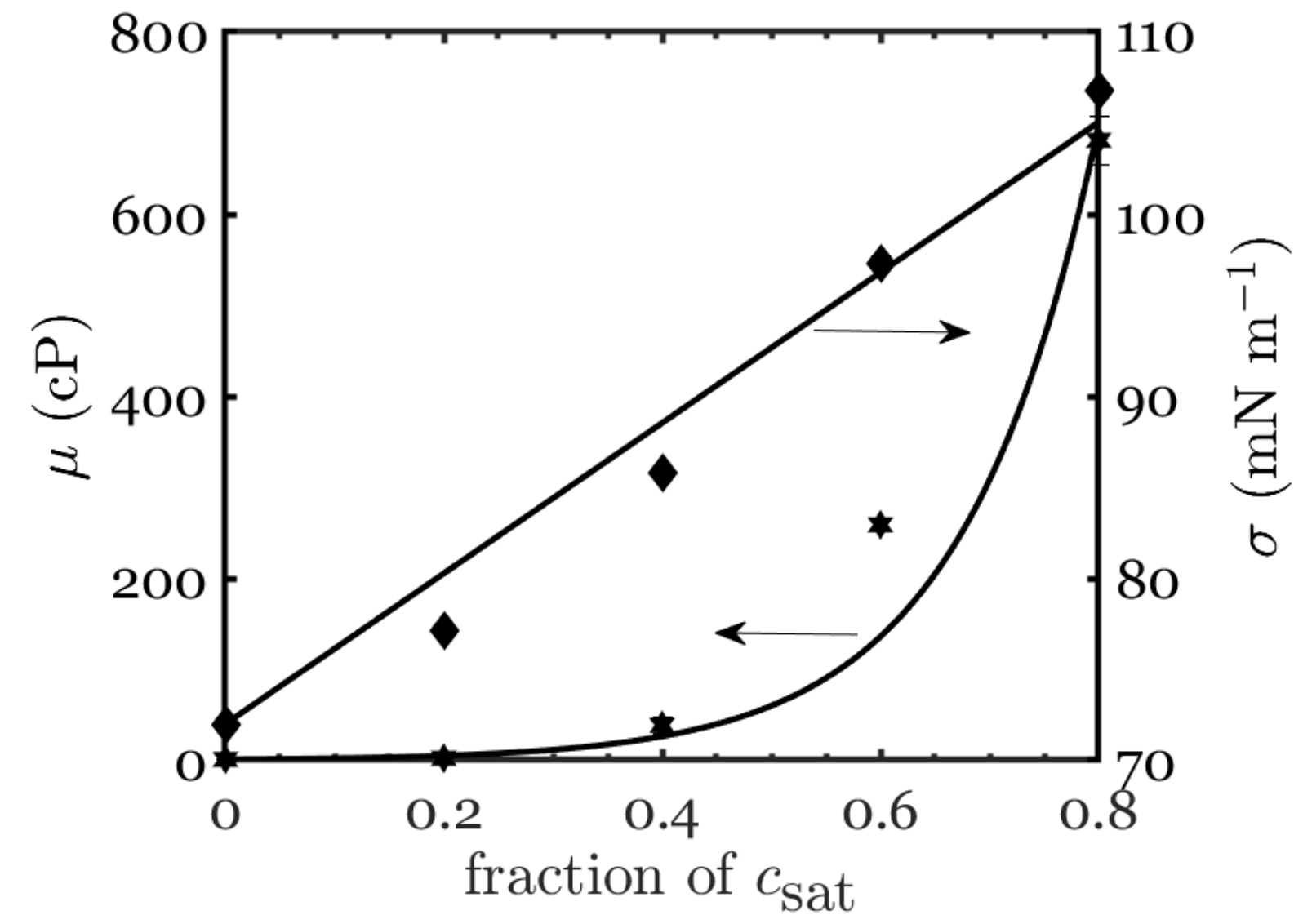

**Figure 3**: Viscosity $\mu$ (plot symbol: hexagram) and surface tension $\sigma$ (plot symbol: diamond) of electrolyte solutions formed by dissolving phosphate salts (sodium phosphate dibasic and potassium phosphate monobasic, 1:1 w/w) in water at concentrations ranging from 0 to 0.8 $c_{sat}$, where $c_{sat} = 7000 \text{ mol m}^{-3}$.

Figure 3 also shows that the dynamic viscosity increases exponentially with phosphate salt concentration in the aqueous medium. This increase can be attributed to the ability of hydrogen phosphate salts to participate in hydrogen bonding, which is known to enhance the viscosity of the medium [19,53]. The viscosity follows an exponential relationship with concentration $\mu = \mu_0 e^{8.2c/c_{sat}}$ where $\mu_0$ is the viscosity of pure water. This Arrhenius-type relationship is consistent with theory and experiments [54–58].

With these findings, we proceeded to perform bulk foaming experiments to observe the effect of salt concentration on foam stability, comparing the phosphate solutions to a control using deionized water.

**Bulk foam test**

Figure 4a and Movie 1 show the stark contrast in foam stability between pure water and phosphate salt solutions. In pure water, the foam collapses almost immediately after the airflow is stopped. In contrast, the foam in the electrolyte solution remains stable even after the cessation of airflow. The bubble population in the salt solution is significantly larger, with smaller bubble sizes compared to those in pure water (Figure 4a and Movie 1). Remarkably, some bubbles in the salt solution persisted for over an hour, as shown in Figure 4b.

This enhanced foam stability in the presence of salts prompted a deeper investigation into the mechanisms behind this phenomenon. To better understand the stabilization effect, we conducted systematic single-bubble experiments using a dynamic fluid interferometer, as detailed in the next section.

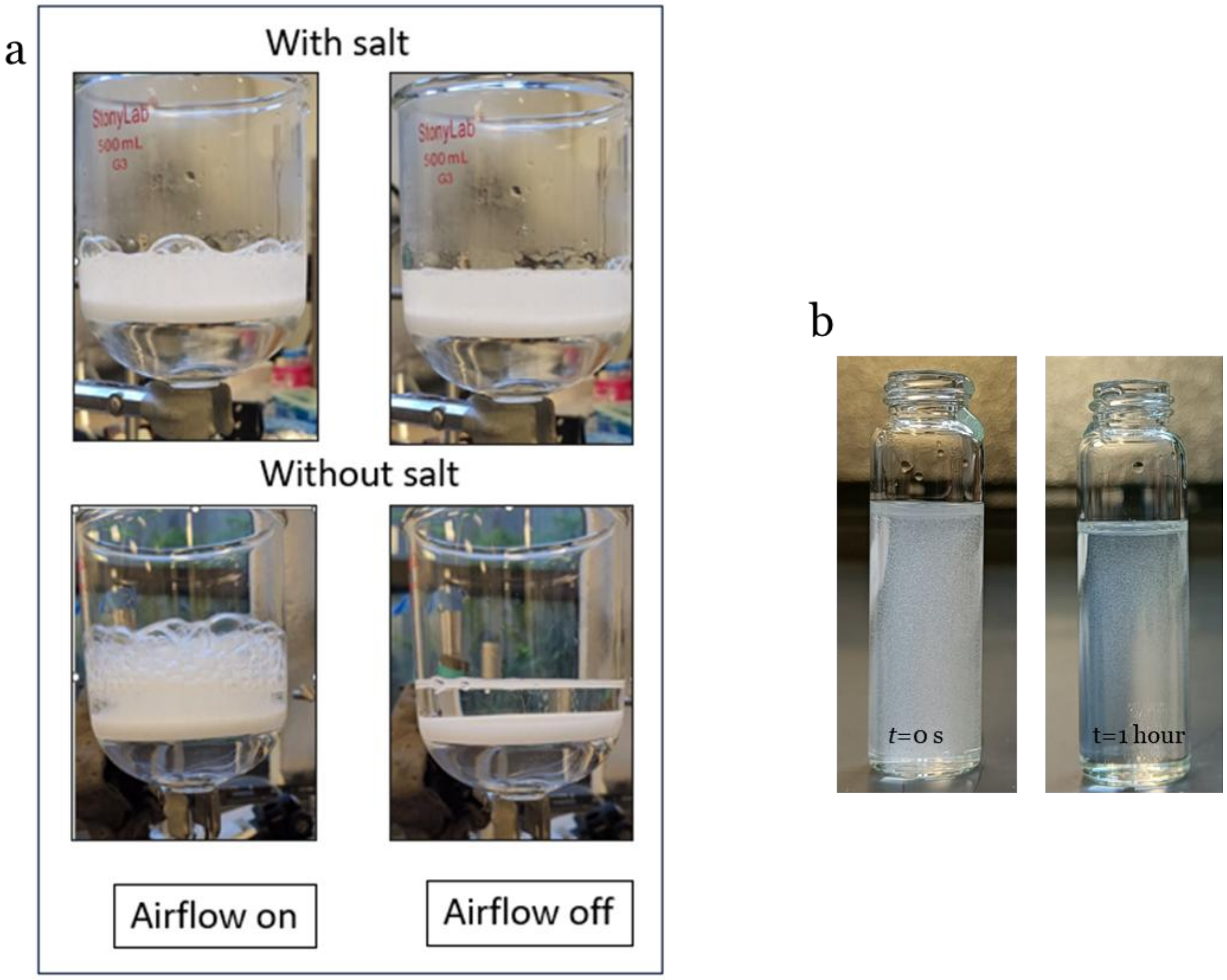

**Figure 4:** (a) Bulk foam stability test comparing a salt solution (0.6 $c_{sat}$ of phosphate salts dissolved in deionized water) and deionized water. (b) A sample of the salt solution transferred from the foaming apparatus to a vial and observed after 1 hour.

### Effect of electrolyte concentration on film drainage dynamics

In the single-bubble interferometric experiments, the flat air-water interface is brought toward a stationary bubble, creating a dimple-shaped film whose drainage dynamics are captured using white light interferometry. As a control experiment, we first conducted dynamic fluid interferometry (DFI) on an air bubble immersed in deionized water, interacting with a flat air-water interface. Figure 5a and Movie 2 show a series of interferograms depicting the behavior of the film in deionized water. We also performed experiments using concentrated phosphate solutions at 20%, 40%, and 60% of their saturation concentration (Figures 5b-d and Movies 3-5). Qualitatively, we observed that the time to film rupture increased with higher salt concentrations. Several noteworthy features emerged in the presence of the electrolytes, with bubble lifetimes extending from a few seconds in deionized water to several minutes in the phosphate solutions. In deionized water, the pressure-induced dimple collapses within a second, whereas it persists for significantly longer in the electrolyte solutions.

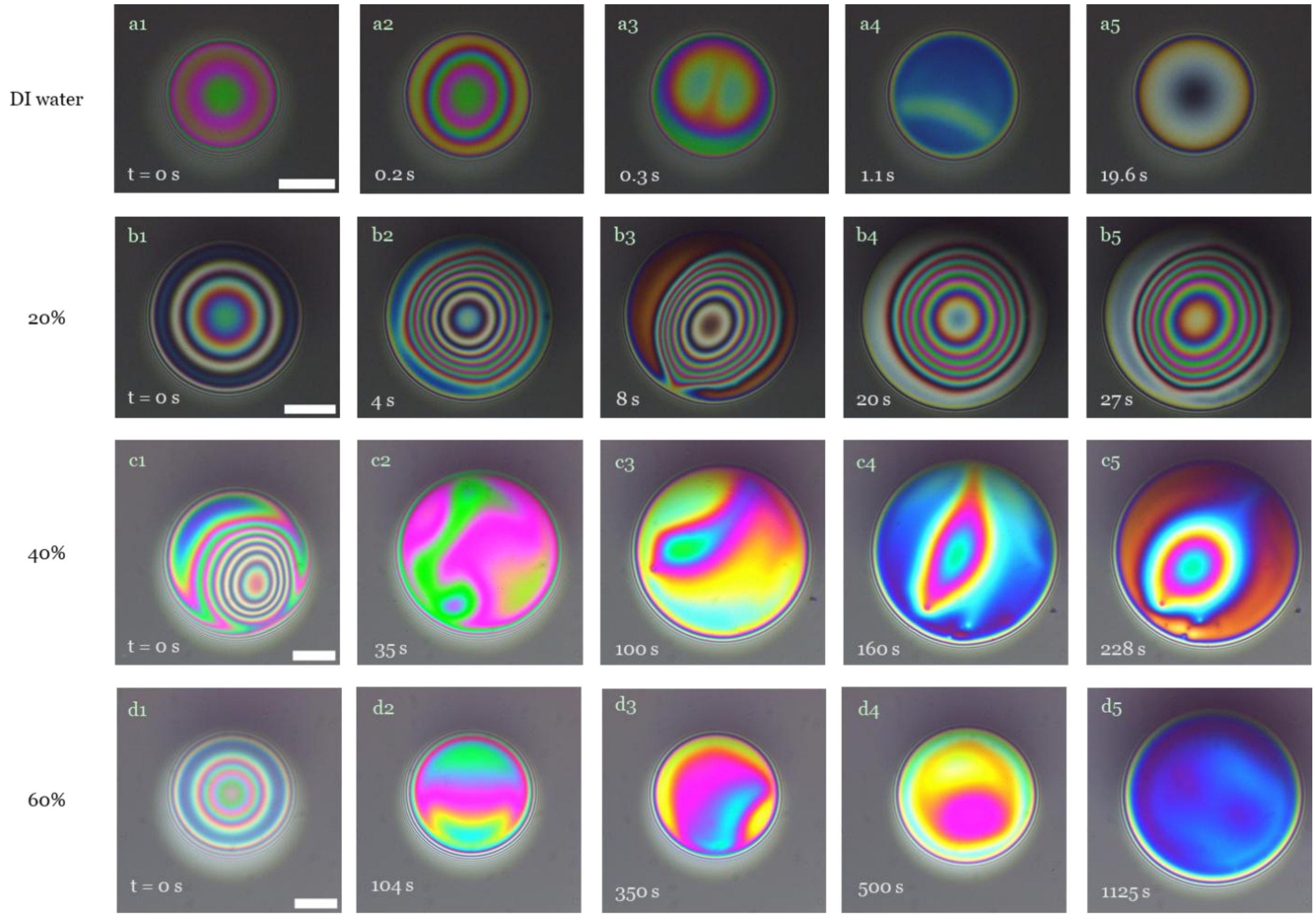

**Figure 5:** Time-lapse images showing foam film evolution over time. (a1-a5) Deionized water, (b1-b5) electrolyte solution with concentration 0.2 $c_{sat}$, (c1-c5) electrolyte solution with concentration 0.4 $c_{sat}$, and (d1-d5) electrolyte solution with concentration 0.6 $c_{sat}$. The image sequences demonstrate the enhanced stability of foam films with increasing concentrations of dissolved phosphate salts. Scale bar: 100 μm.

Using the interferograms, we measured both the minimum and apex heights of the film, which allowed us to describe the dynamics of film drainage [59–61] (Figures 6a and 6b). A clear relationship emerged between the thinning process and the dominant pressure-driven component of velocity, consistent with the drainage of a dimpled film [25] (Figure 6c). By modeling the rate of film thinning through a mass balance approach, we derive the following equation [25]:

$$\frac{dh}{dt} \sim -\frac{h}{R_f} u_p, \qquad (1)$$

where $h$ is the film thickness, $t$ is time, $R_f$ the film radius, and $u_p$ is the parabolic velocity component. Using lubrication theory, the parabolic velocity component can be expressed as:

$$u_{\mathrm{p}} \sim \frac{h^2}{\mu} \frac{p}{\ell_{\mathrm{c}}}, \tag{2}$$

where $p$ is the lubrication pressure, and $\ell_c$ is the characteristic length scale of the pressure gradient. For a dimpled film, the lubrication pressure corresponds to the capillary pressure $(\sigma / R)$ [25,26]. The characteristic length scale, $\ell_{\mathrm{c}}$, can be shown to be $\sqrt{hR}$ for both spherical and dimpled films [62]. Substituting these into equation (1) and integrating from $t = 0$ to $t$, we obtain:

$$\frac{h}{h_0} = \left(1 + \alpha \frac{\sigma h_0^{3/2}}{\mu R_{\mathrm{f}} R^{3/2}} t\right)^{-2/3} = \left(1 + \beta t\right)^{-2/3}, \tag{3}$$

where $h(t = 0) = h_0$, $\alpha$ is a scalar constant, and $\beta = \alpha \sigma h_0^{3/2} / \mu R_{\mathrm{f}} R^{3/2}$.

In our experiments, $\alpha$ was found to be 0.105. Additionally, the apex film height was found to be linearly related to the minimum film height, indicating that the dimple height is insignificant, and the change in film volume is mainly governed by the disc portion of the film (Figures 6d and 6e).

In foaming systems, the foam films formed by the topmost air bubbles at the free air-water interface may be subject to evaporation, depending on bubble size, while the films formed between interior bubbles are generally less susceptible to evaporation. In the next section, we demonstrate that our experiments are designed to isolate capillary drainage as the dominant factor in film thinning. Subsequently, we will show that, despite capillary drainage dominance, Marangoni-driven stabilization of the foam film is still observed.

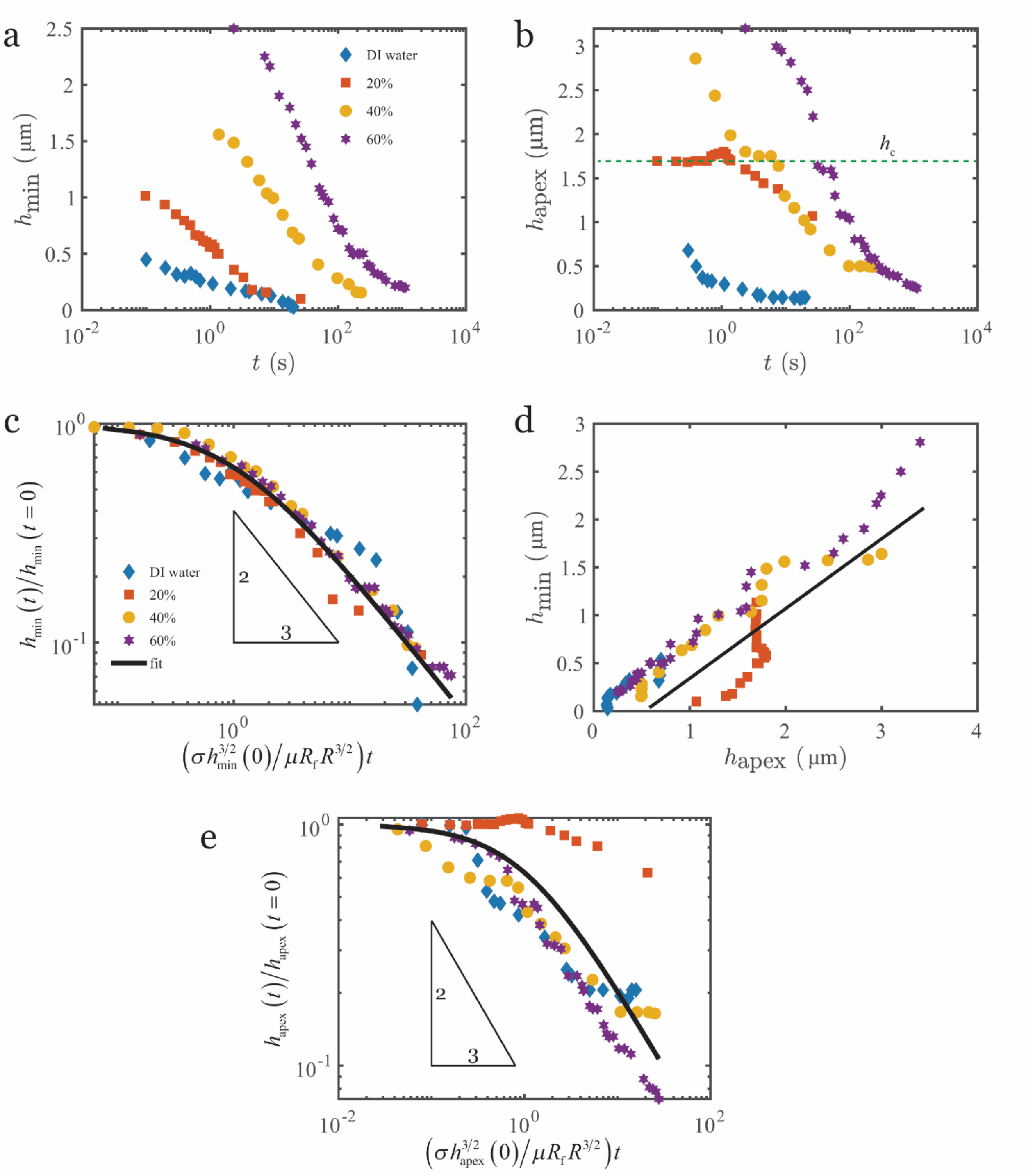

**Figure 6:** (a) Minimum film thickness $(h_{min})$ as a function of time for aqueous solutions containing dissolved phosphates at different concentrations. (b) Film thickness at the apex $(h_{apex})$ as a function of time for aqueous

solutions containing dissolved phosphates at different concentrations. The green dashed line shows the critical height, $h_c$, at which film volume increases. (c) Non-dimensionalized minimum film height as a function of non-dimensionalized time showing a $h_{\mathrm{min}} \sim t^{-2/3}$ relationship suggesting that the film thinning behavior is capillary drainage-dominated. (d) $h_{\mathrm{min}}$ versus $h_{\mathrm{apex}}$ shows a linear relationship suggesting that the film volume is mainly governed by volume of the disc of thickness $h_{\mathrm{min}}$ and radius $R_{\mathrm{f}}$ , and that the dimple height is minimal. (e) Non-dimensionalized apex film thickness as a function of non-dimensionalized time, also showing a $h_{\mathrm{apex}} \sim t^{-2/3}$ relationship. In all the plots, diamonds, square, circle and hexagram symbols represent de-ionized water, 20%, 40%, 60% of $c_{\mathrm{sat}}$ phosphate solutions respectively.

**Film thinning behavior: Hydrodynamic dominance**

The observation that film thinning is dominated by hydrodynamic effects is not surprising and can be further demonstrated using scaling analysis. This can be done by comparing the rates of film thinning due to evaporation and capillary drainage as a function of film thickness for pure water. The rate of film thinning by evaporation can be expressed as $\left(-dh/dt\right)_{\mathrm{evap}} = \left(P_{\mathrm{ve}} - P_{\mathrm{v}}\right) K_{\mathrm{g}} / \rho$ , where $P_{\mathrm{ve}}$ is the equilibrium water vapor pressure, $P_{\mathrm{v}} = R_{\mathrm{H}} P_{\mathrm{ve}}$ is the actual water vapor pressure, $R_{\mathrm{H}}$ is the relative humidity, and $K_{\mathrm{g}} = DM_{\mathrm{w}} / k_{\mathrm{B}} N_{\mathrm{A}} T \delta$ is the mass transfer coefficient, $D$ is the diffusivity of water vapor in air, $k_{\mathrm{B}}$ is the Boltzmann constant, $N_{\mathrm{A}}$ is the Avogadro number, $T$ is the absolute temperature, and $\delta$ is the boundary layer thickness, which equals the film radius $R_{\mathrm{f}}$ [19,63,64].

In contrast, the rate of film thinning due to capillary drainage is given by $\left(dh/dt\right)_{\mathrm{cap}} = \left(\sigma h^{5/2}\right) / \left(\mu R^{3/2} R_{\mathrm{f}}\right)$ where $\sigma$ is the surface tension, $h$ is the film thickness, $\mu$ is the viscosity, $R$ is the bubble radius, and $R_{\mathrm{f}}$ is the film radius.

To assess the relative contribution of these two mechanisms, we define a parameter, $\zeta = \left(-dh/dt\right)_{\mathrm{evap}} / \left(-dh/dt\right)_{\mathrm{cap}}$ as the ratio of the evaporation rate to the capillary drainage rate. By plotting $\zeta$ as a function as film thickness $h$, we find that $\zeta < 1$ until the film thickness reaches approximately 100 nm, indicating that capillary drainage dominates over evaporation for the majority of the film thinning process (see Figure 7).

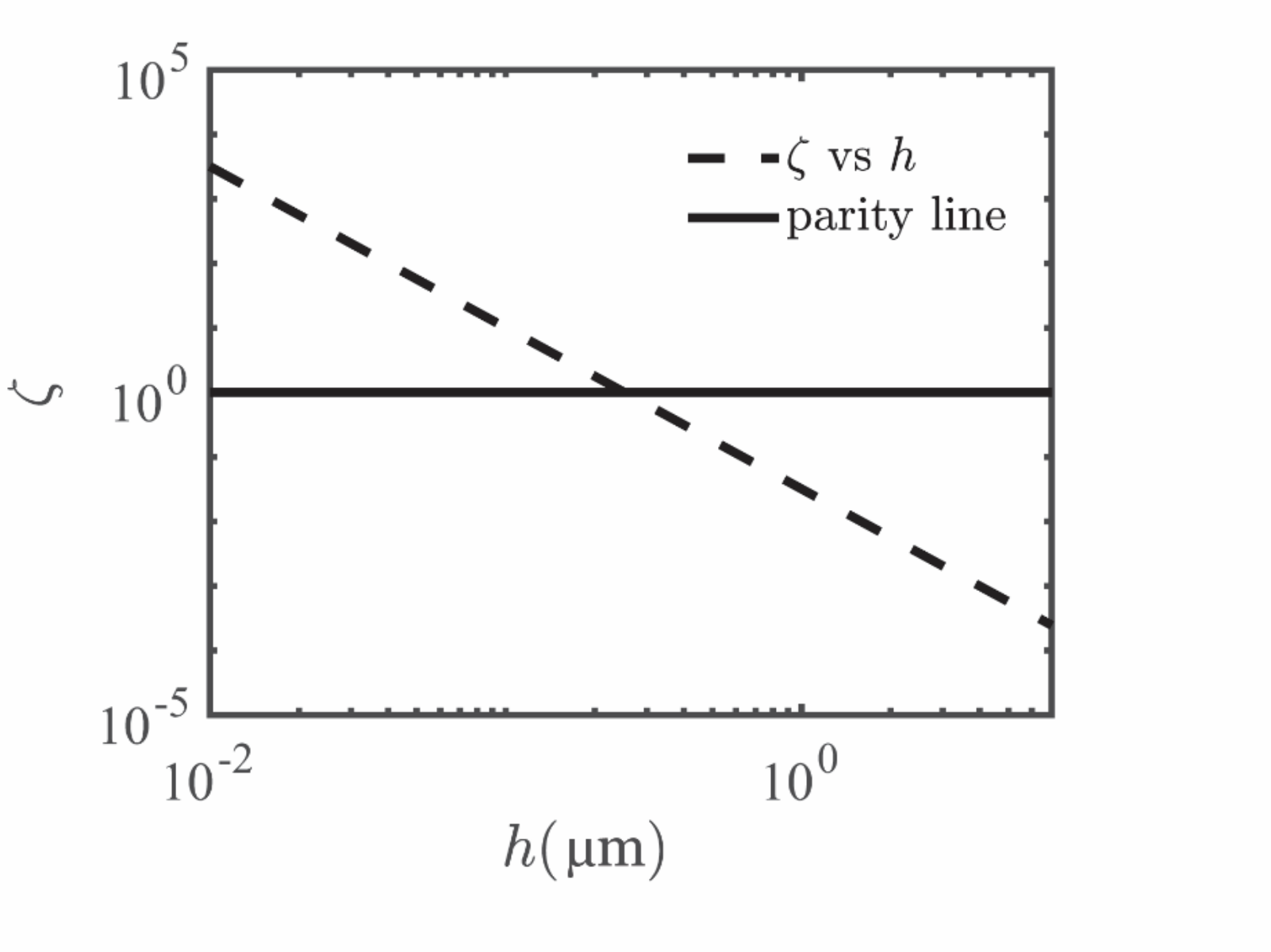

**Figure 7:** Plot of the parameter $\zeta = \left(-dh/dt\right)_{\text{evap}} / \left(-dh/dt\right)_{\text{cap}}$ versus film thickness, showing that film thinning is expected to be capillary-dominated $\left(\zeta < 1\right)$ for the majority of the thinning process.

Given that the temporal behavior of the film thickness follows hydrodynamic drainage, one might attribute the enhanced film stability solely to the increased viscosity of the bulk fluid. Additionally, higher salt concentrations lead to thicker films during the initial stages of drainage, as shown in Figure 3. This increase in film thickness can be explained by the elevated viscosity of the salt solutions at higher concentrations. As the flat air-water interface approached the stationary bubble at a constant velocity, more force is required to displace the more viscous solution. The force acting on the film can be expressed as $\mu G R^2$ where $G$ is the strain rate.

## Intriguing observations and comparison with existing models

A striking observation arises from the apex thickness $\left(h_{\text{apex}}\right)$, which exhibits a distinctive kink at approximately the same thickness (~1.7 μm) for all three salt concentrations (see Figures 6b and 8). This is surprising, given that the film drainage dynamics are clearly dominated by hydrodynamics, not evaporation.

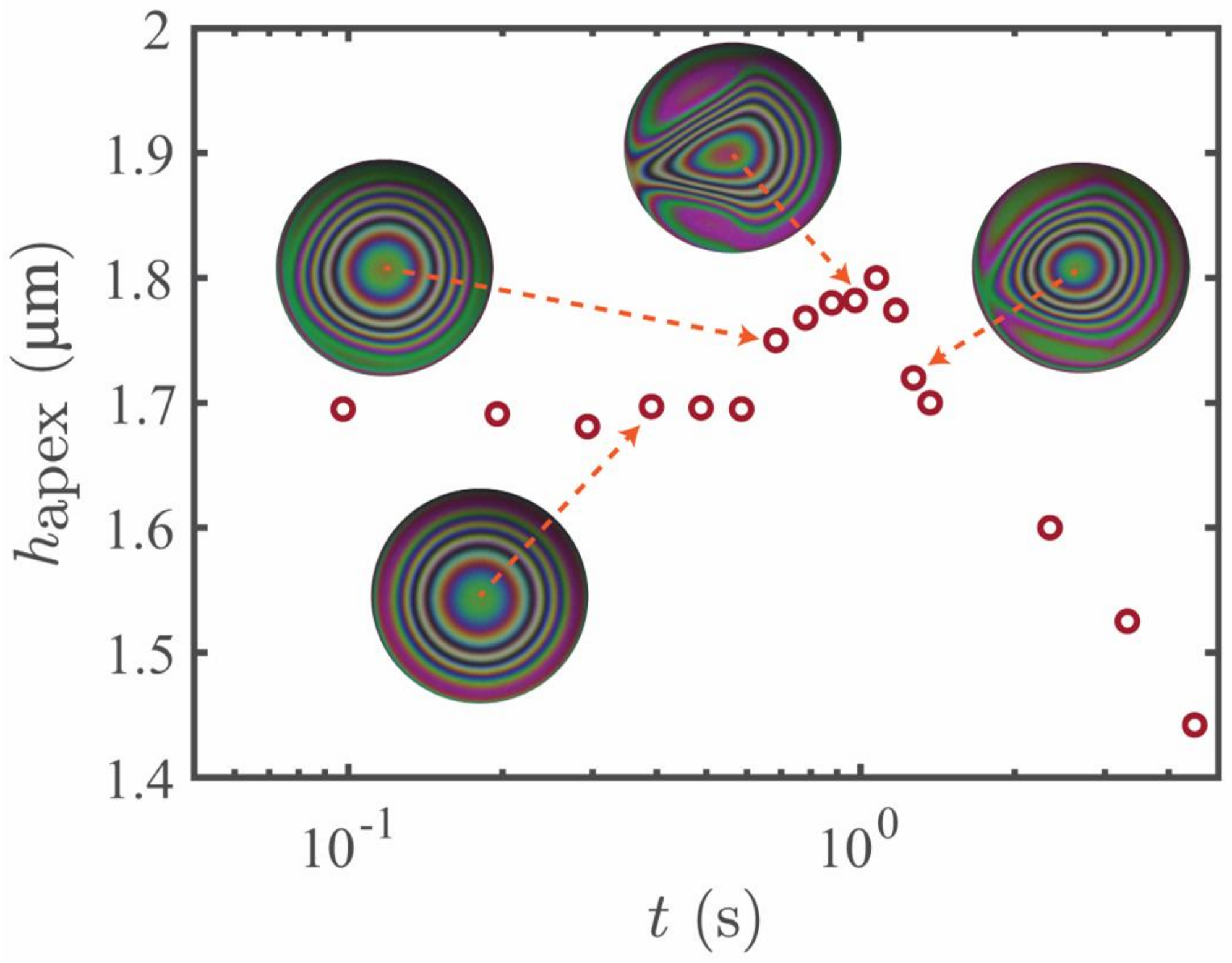

**Figure 8:** Increase in apex thickness caused by Marangoni flows in a salt solution (concentration = 0.2 $c_{\text{sat}}$). Based on experimental data, the expected surface tension difference between the film region and the bulk fluid is $\Delta\sigma = 0.1\ \text{mN m}^{-1}$, which aligns with predictions from Marrucci's theory.

This prompted us to investigate the existing literature for other possible explanations. Recently, Liu and co-workers proposed a mechanism for delayed coalescence of air bubbles immersed in low-concentration (~100 mM) salt solutions, such as NaCl, $CaCl_2$, $Mg(ClO_4)_2$, NaOH, and $HNO_3$ [45]. Their model assumes a mobile air-water interface and ion depletion near the interface. Under these conditions, they derived an electrolyte transport equation, expressing the change in electrolyte concentration $C$ in the film as $DC/Dt = -\left[ h_{\text{salt}} / \left( h + h_{\text{salt}} \right) \right] \left( C/h \right) \left( Dh/Dt \right)$ where $h_{\text{salt}} = -2\left[ 1/k_{\text{B}} N_{\text{A}} T \left( 1 + \varepsilon \right) \right] \left( d\sigma / dC \right)$, $d\sigma / dC$ representing the variation of surface tension with concentration, and $\varepsilon$ a parameter accounting for the non-ideality of the solution mixture.

In their model, $h_{\text{salt}}$ is approximately 2 nm, and $DC/Dt$ becomes significant only when the film thickness $h$ approaches $h_{\text{salt}}$. This leads to the conclusion that film drainage is arrested when $h \sim 30\ \text{nm}$.

However, this model does not explain the slowdown or reversal of film drainage at larger film thicknesses, such as those observed in our experiments where $h \sim 1.5\ \mu\text{m}$. Similar large critical thicknesses have been reported by Karakashev and co-workers [46]. Additionally, Liu's model is based on highly controlled conditions, where interfaces are carefully maintained as mobile. In most practical scenarios, however, interfaces become immobile due to trace contaminants [65,66], as observed in our experiments. The film drainage dynamics in our case match a dimple-shaped, parabolic behavior with no-slip conditions at both interfaces [25] [see equation (3)].

Importantly, the assumption of a mobile interface implies the presence of salts at the air-fluid interface. However, recent experimental findings suggest a more nuanced interfacial structure: the topmost air-water layer is devoid of solutes, while an intermediate sub-surface layer, located a few angstroms below, contains solutes [24]. This stratification allows for solute motion even in parabolic flow-dominated film drainage, where the interfacial velocity is zero. These findings reconcile the observed Marangoni-driven effects with immobile interface conditions, providing a new perspective on film drainage dynamics.

We now turn to the seminal paper by G. Marrucci [12], which we believe has been prematurely dismissed in the past due to less precise experimental evidence compared to modern interferometric techniques. Marrucci's theory posits that as a thin film drains, a control volume $V$ within the film experiences an increase in interfacial area $(\Delta s)$ (see Figure 9). If $c_i$ and $\Gamma_i$ represent the bulk concentration and Gibbs surface excess of species 'i' in the volume $V$ with interfacial area $s$, then as the film thins and the interfacial area increases to $s + \Delta s$, a mass balance on species 'i' can be written as:

$$c_i V + \Gamma_i s = (c_i + \Delta c_i)V + (\Gamma_i + \Delta\Gamma_i)(s + \Delta s). \quad (4)$$

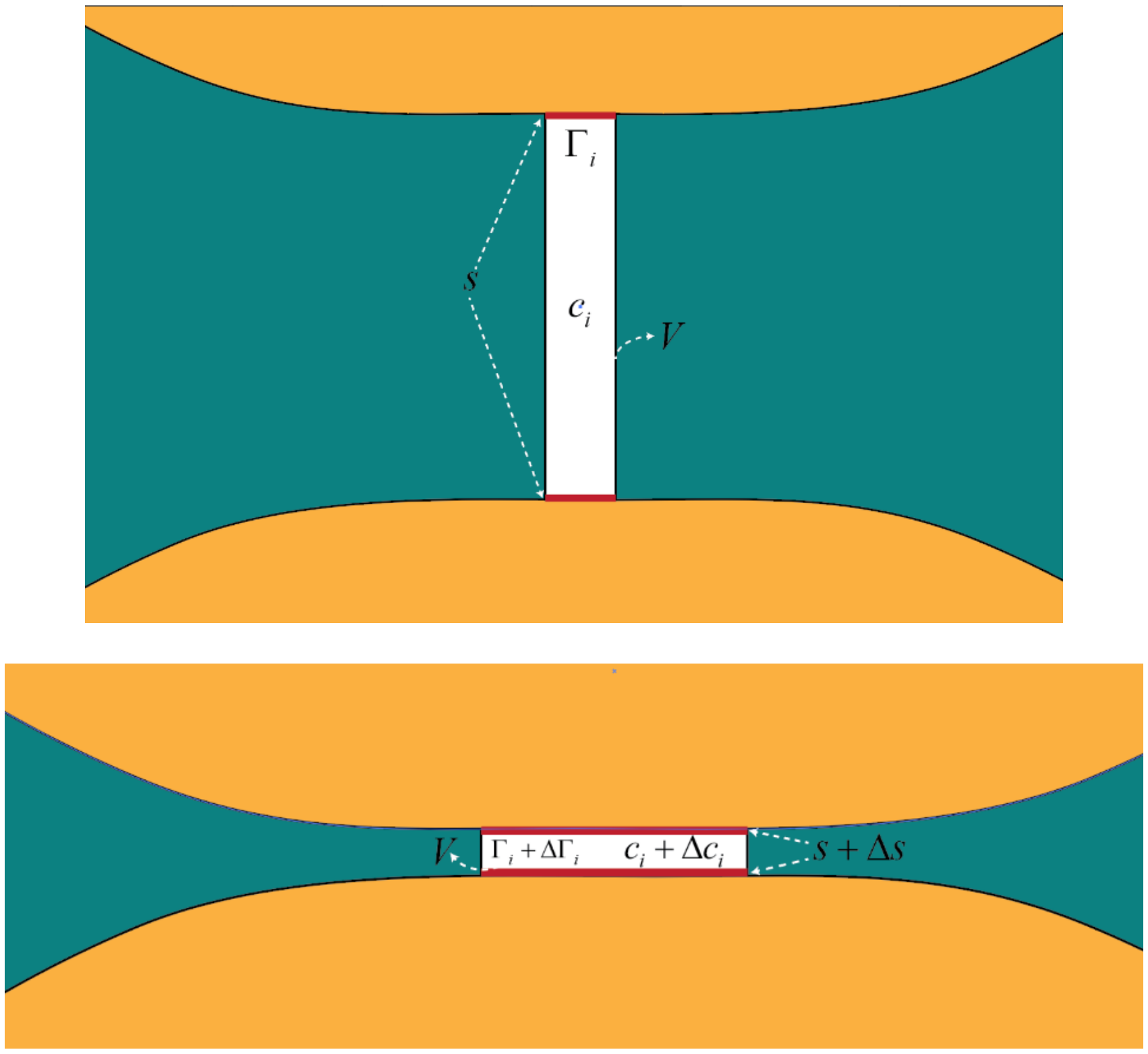

**Figure 9**: Schematic illustrating that as the film thins, the interfacial area experienced by the control volume $V$ increases by an amount $\Delta s$. This increase in interfacial area leads to the rearrangement and mass transfer of species such that the Gibbs surface excess reaches equilibrium $(\Delta\Gamma_i = 0)$. Through a mass balance and assuming high Péclet numbers, it can be shown that $\Delta c_i \neq 0$. Specifically, $\Delta c_i > 0$ if species 'i' has a higher surface tension, and vice versa. This results in a Marangoni-driven influx and stabilization of the film in a multi-component mixture.

By simplifying the equation and assuming $\Delta\Gamma_i = 0$ [12], since mass transfer rates are higher in nanometric films, leading to faster equilibration of surface coverage, we obtain the change in the bulk concentration of species 'i' in the volume $V$ as:

$$\Delta c_i = -\Gamma_i \frac{\Delta s}{V} \approx -\frac{\Gamma_i}{h} \quad (5)$$

For small changes in concentration, we can approximate the change in surface tension for a mixture of two species as:

$$\Delta\sigma = \frac{d\sigma}{dc_1}\Delta c_1 + \frac{d\sigma}{dc_2}\Delta c_2 \tag{6}$$

Equation (6) is important because it shows that surface tension in the film will always increase as the film thins, provided the Marangoni flow-based Péclet number is sufficiently high. If species 1 has a lower surface tension, then $\Gamma_1 > 0$ and $\Gamma_2 < 0$, with $d\sigma / dc_1 < 0$ and $d\sigma / dc_2 > 0$. Consequently, from equations (5) and (6), $\Delta\sigma > 0$.

Using the definition of $\Gamma_i$ which represents the excess number of moles of species 'i' in the interfacial region relative to the bulk, we obtain the relationship $\sum \Gamma_i v_i = 0$, where $v_i$ is the molar volume of species 'i' [67]. From the Gibbs-Duhem relationship, $\sum x_i d\ln c_i = 0$. Additionally, the thermodynamic relationship between surface tension $\sigma$ and $\Gamma_i$ is given by $-d\sigma = k_B T \sum d\ln c_i$ [67]. Using these relationships, one can derive the Gibbs surface excess for a two-component mixture when the mole fraction of species 1 is much smaller than that of species 2:

$$\Gamma_1 = -\frac{c_1}{N_A k_B T}\frac{d\sigma}{dc_1}. \tag{7}$$

From equations (5), (6) and (7), we get the surface tension difference is given by:

$$\Delta\sigma = \frac{2c_1}{N_A k_B T h}\left(\frac{d\sigma}{dc_1}\right)^2. \tag{8}$$

The surface tension difference produces a Gibbs-Marangoni pressure [12] , $p_{GM}$ , expressed as

$$p_{\mathrm{GM}} = \frac{2\Delta\sigma}{h} = \frac{4c_1}{N_{\mathrm{A}} k_{\mathrm{B}} T h^2}\left(\frac{d\sigma}{dc_1}\right)^2. \tag{9}$$

When the Gibbs-Marangoni pressure scales with the Laplace pressure $(\sigma / R_{\mathrm{H}})$, where $R_{\mathrm{H}}$ is the effective radius given by the harmonic mean, we can estimate the critical film height, $h_{\mathrm{c}}$, at which flow reversal is expected. This balance yields $h_{\mathrm{c}}$ as:

$$h_{\mathrm{c}} = \frac{d\sigma}{dc_1}\sqrt{\frac{4c_1 R_{\mathrm{H}}}{\sigma N_{\mathrm{A}} k_{\mathrm{B}} T}}, \tag{10}$$

which can be found in Marrucci's work [12]. The parameter $R_{\mathrm{H}}$ the effective radius of curvature, which is defined based on the drainage geometry. For film drainage occurring between two curved fluid interfaces with radius of curvature $R$, $Rf=R/2$R f =R/2. For drainage between a curved interface of radius $R$R and a flat interface with infinitely large curvature, $Rf=R$R f =R.

To test Marrucci's theory, we first compute the expected surface tension difference produced for concentration $c = 0.2c_{\mathrm{sat}} = 1400\ \mathrm{mol\ m^{-3}}$.

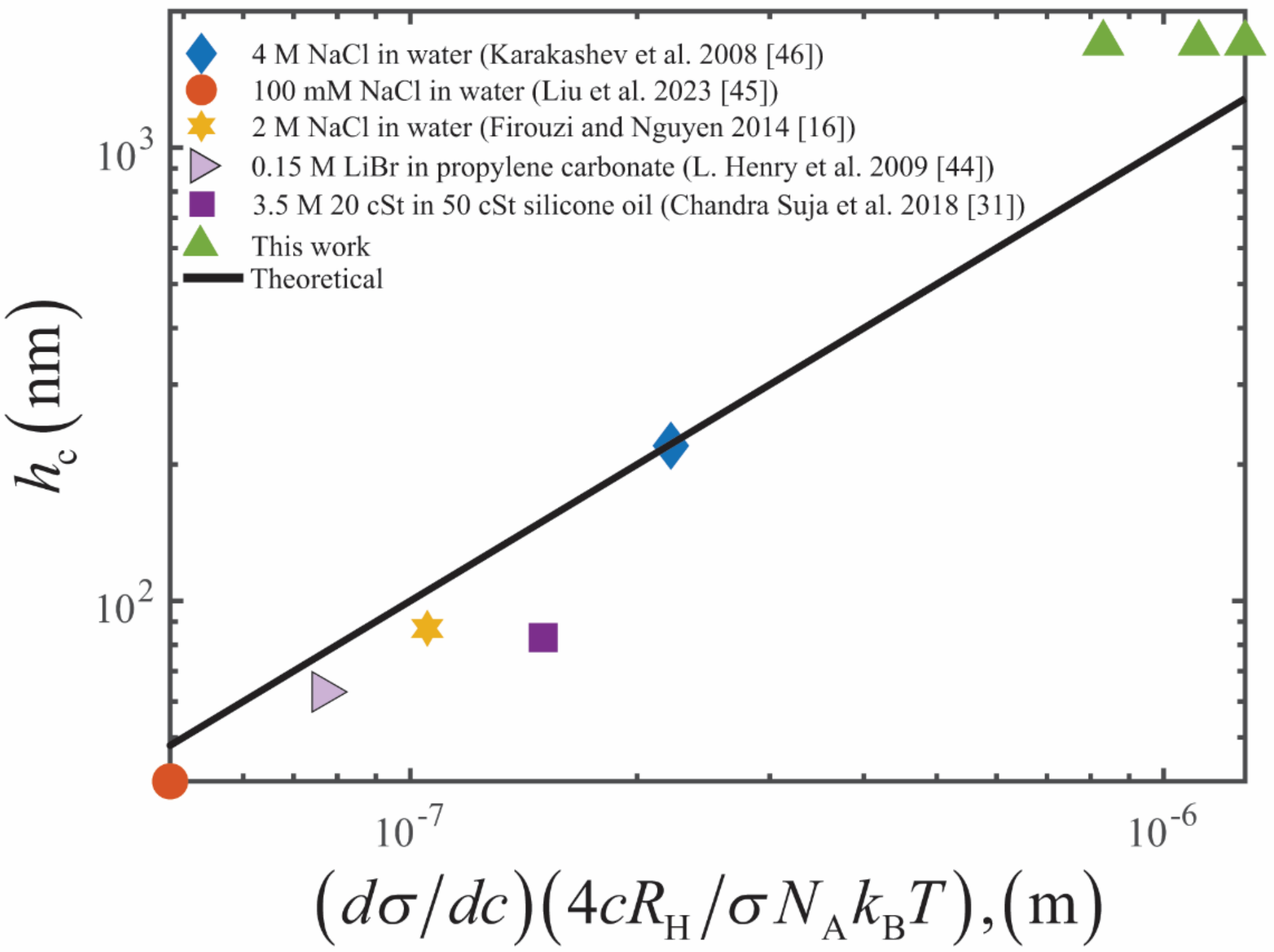

Figure 10: Plot showing that the critical height $h_c$ predicted by Marrucci's theory agrees well with our data for phosphate salts in aqueous media, as well as NaCl at varied concentrations [45–47], and foam films in non-aqueous media such as silicone oil blends [31,41–44]. The parameter $R_H$ is the effective radius of curvature, which is defined based on the drainage geometry. For film drainage occurring between two curved fluid interfaces with radius of curvature $R$, $R_H = R/2$. For drainage between a curved interface of radius $R$ and a flat interface with infinitely large curvature, $R_H = R$.

Taking $d\sigma / dc = \left(44.3\times10^{-3} / 7000\right)$ $\mathrm{Nm^3mol^{-1}} = 6.3\times10^{-6}$ $\mathrm{Nm^3mol^{-1}}$ and $h = 1.5$ μm, we get $\Delta\sigma = O(0.1)$ mN $\mathrm{m^{-1}}$ (see equation 10). Using data in Figure 8, we estimate a film thickening rate $\left(dh / dt\right) = (1.8 - 1.5) / (1 - 0.6)$ μm $\mathrm{s^{-1}} = 0.7$ μm $\mathrm{s^{-1}}$, and the radial influx velocity, $u_{\mathrm{influx}} = \left[\left(dh / dt\right)_{\mathrm{influx}} R_f\right] / h = 0.9$ μm $\mathrm{s^{-1}}$, where the film radius, $R_f = 180$ μm. Using the scaling relationship, $u_{\mathrm{influx}} = h\Delta\sigma_{\mathrm{exp}} / \mu R_f$ and viscosity at $c = 0.2c_{\mathrm{sat}}$, $\mu \sim 10$ cP (see Figure 3), we can estimate $\Delta\sigma_{\mathrm{exp}} \approx 0.1$ mN $\mathrm{m^{-1}}$, suggesting that Marrucci's theory is valid.

We tested Marrucci's prediction [equation(10)] using experimental data from three previous studies that used interferometry to investigate the stability of air bubbles in aqueous NaCl solutions [45–47]. These studies utilized NaCl concentrations of 100 mM, 2 M, and 4 M. Using the known variation of surface tension with concentration for NaCl $\left(d\sigma/dc\right)$, we calculated the critical film height $h_c$ using equation(10). These predicted critical heights were then compared with the experimentally measured values from the studies (see Figure 10 and Table 1).

As shown in Figure 10 and Table 1, equation (10) accurately predicts the critical heights, demonstrating good agreement with Marrucci's theory and the experimental data.

We further tested the applicability of equation (10) to foam stability in non-aqueous media. One such example comes from the work of Chandra Suja et al. (2018), where small amounts of low molecular weight, high volatility, and low surface tension silicone oils were blended with high molecular weight, low volatility, and high surface tension silicone oils [31]. In experiments where the system was exposed to evaporation, large oscillations in film thickness were observed and attributed to Marangoni stresses, driven by significant surface tension differences due to the depletion of low molecular weight oils in the film region. However, even when evaporation was suppressed by sealing the chamber, unusually stable foam films (~100 nm) were observed—an aspect that was not fully addressed by the authors.

By applying Marrucci's equation (10) to this system, we obtained critical heights that were in very good agreement with the experimentally observed stable films, even in a closed system (see Figure 10 and Table 1).

This finding provides a profound insight: while the stability of the top layers of bubbles in bulk foam may be influenced by evaporation and Marangoni-driven stabilization, the stability of bubbles deeper within the foam, where evaporation is not a factor, may be explained solely by Marrucci's theory, as captured in equation (10).

**Table 1:** Comparison between critical heights $\left(h_c\right)$ calculated using Marrucci's theory and experimental observations from this study and previous literature that employed interferometric techniques in film drainage studies. The table also includes the parameters used to compute the critical heights based on Marrucci's theory.

| Study | Salt type and concentration | $\frac{d\sigma}{dc}\times10^6$ $\left(\mathrm{Nm^3\ m^{-1}\ mol^{-1}}\right)$ | Bubble radius, $R$ $(\mu\mathrm{m})$ | Surface tension, $\sigma$ $\left(\mathrm{mN\ m^{-1}}\right)$ | $h_c\,(\mathrm{nm})$, Marrucci's theory | $h_c\,(\mathrm{nm})$, experimental |
|---|---|---|---|---|---|---|
| Karakashev et al. 2008 **[46]** | NaCl in water, 4M | 1.85 | 350 | 80 | 265 | 220 |
| Liu et al. 2023 **[45]** | NaCl in water, 100mM | 1.85 | 590 | 72 | 48 | 30-50 |
| Firouzi and Nguyen 2014 **[47]** | NaCl in water, 2 M | 1.85 | 150 | 76 | 105 | 87 |
| Chandra Suja et al. 2018 **[31]** | 20 cSt Silicone oil in 50 cSt silicone oil, 3.45 mM | 9.05 | 1000 | 20.6 | 150 | 83 |
| L. Henry et al. 2009 **[44]** | LiBr in propylene carbonate, 0.15 M | 0.996 [68] | 2000 | 41.488 [68] | 77 | 63 |
| This study | Phosphate salts in water, 1.4M to 4.2M | 6.34 | 576 | 77 to 97 | 830 to 1282 | 1700 |

## Conclusions

In this study, we revisited the long-standing question of foam stabilization in electrolyte solutions, focusing on the effects of electrolyte concentration on film drainage dynamics. Additionally, we demonstrated that this stabilization mechanism applies to non-aqueous foams as well. Through precise interferometric single-bubble experiments, we demonstrated that Marrucci's 1969 theory offers a compelling explanation for the observed phenomena. Marrucci's theory posits that a control volume within the film experiences an increase in interfacial area as the film thickness decreases. When the Marangoni-based Péclet number is high, the surface concentrations of all species in the film reach equilibrium due to the high mass transfer rates in nanometric films. However, the bulk concentration of the species associated with higher surface tension increases in the film region, while that of lower surface tension species decreases. This creates concentration gradients that lead to surface tension gradients and ultimately a Marangoni-driven influx.

Our results showed that capillary drainage dominates the film thinning process, due to the smaller film radii produced during our experiments. Nevertheless, we observed an anomalous influx and an increase in film volume, the magnitude of which closely matched the predictions of Marrucci's theory.

We validated Marrucci's theory by calculating critical film heights for a range of electrolyte concentrations, including NaCl solutions from previous studies, and comparing them with experimental data [43-45]. Marrucci's theory consistently predicted critical heights that closely matched experimental observations. This validation extended beyond aqueous systems; in non-aqueous media, such as silicone oil blends, the theory also accurately predicted foam film stabilization in closed systems, where evaporation was suppressed, further demonstrating its wide applicability [30].

Our findings suggest that Marrucci's theory provides critical insights into the stability of bubbles within bulk foams, particularly for bubbles situated below the surface layers, where evaporation is minimal.

In conclusion, this work not only confirms the validity of Marrucci's theory in predicting critical heights and foam stability but also highlights its broad applicability across different systems. Our study reinforces the importance of considering the interplay between hydrodynamic and evaporation effects in foam stabilization, offering new avenues for the design and optimization of foams in industrial applications.

## Acknowledgements

Author DM was supported by departmental core grants from the NIH (P30 EY026877) and Research to Prevent Blindness.